\documentclass[aps,prb,twocolumn,showpacs,superscriptaddress,groupedaddress]{revtex4}  

\usepackage{graphicx, epstopdf}  
\usepackage{dcolumn}   
\usepackage{bm}        
\usepackage{amssymb,amsmath}   
\usepackage{color,epsfig,float}
\DeclareMathOperator{\Tr}{Tr}
\begin{document}


\title{Control of spin dynamics in a two-dimensional electron gas by electromagnetic dressing}

\author{A. A. Pervishko}
\affiliation{Division of Physics and Applied Physics, Nanyang
Technological University 637371,  Singapore}

\author{O. V. Kibis}\email{Oleg.Kibis@nstu.ru}
\affiliation{Department of Applied and Theoretical Physics,
Novosibirsk State Technical University, Novosibirsk 630073,
Russia} \affiliation{Division of Physics and Applied Physics,
Nanyang Technological University 637371,  Singapore}

\author{S. Morina}
\affiliation{Division of Physics and Applied Physics, Nanyang
Technological University 637371,  Singapore} \affiliation{Science
Institute, University of Iceland  IS-107, Reykjavik, Iceland}

\author{I. A. Shelykh}
\affiliation{Division of Physics and Applied Physics, Nanyang
Technological University 637371,  Singapore} \affiliation{Science
Institute, University of Iceland IS-107, Reykjavik, Iceland}
\affiliation{ITMO University, St. Petersburg 197101, Russia}


\begin{abstract}

We solved the Schr\"odinger problem for a two-dimensional electron
gas (2DEG) with the Rashba spin-orbit interaction in the presence
of a strong high-frequency electromagnetic field (dressing field).
The found eigenfunctions and eigenenergies of the problem are used
to describe the spin dynamics of the dressed 2DEG within the
formalism of the density matrix response function. Solving the
equations of spin dynamics, we show that the dressing field can
switch the spin relaxation in the 2DEG between the cases
corresponding to the known Elliott-Yafet and D'yakonov-Perel'
regimes. As a result, the spin properties of the 2DEG can be tuned
by a high-frequency electromagnetic field. The present effect
opens an unexplored way for controlling the spin with light and,
therefore, forms the physical prerequisites for creating
light-tuned spintronics devices.

\end{abstract}

 \maketitle

\section{Introduction} One of the most exciting trends in
modern condensed matter physics is using the electron
spin-of-freedom to store and transfer information. This field of
research --- which is known as spintronics --- opened a way for
various high-performance devices which have a number of important
advantages as compared to conventional electronics, including
growth in data processing speed, reduction in power consumption,
etc \cite{Wolf, Awschalom, Zutic,  Bader, Hirohata,Awschalom1}.
Besides successful spintronic experiments based on various
ferromagnetic structures \cite{Ohno1, Lou, Nasirpouri}, an
alternative approach to use nonmagnetic semiconductor
nanostructures with spin-orbit interaction is actively
investigated in recent years \cite{Kato, Awschalom2, Dietl}.
Therefore, the study of spin transport in a two-dimensional
electron gas (2DEG) with the spin-orbit interaction is currently
in the focus of attention. One of the most important
characteristics of spintronics devices is the spin relaxation time
which describes the spin evolution. Since it is responsible for
the spin transfer of information, the search of ways to control
this time is interesting from both fundamental and applied
viewpoint. In the present paper, we report a novel method to
control the spin relaxation time of 2DEG with a strong
high-frequency electromagnetic field.

It is well-known that the interaction between electrons and a
strong high-frequency electromagnetic field cannot be described as
a weak perturbation. In this case, the system ``electron +
electromagnetic field'' should be considered as a whole. Such a
bound electron-field system, which was called ``electron dressed
by field'' (dressed electron), became a commonly used model in
modern physics \cite{Scully, Cohen-Tannoudji}. Recently, we
demonstrated that strong interaction between 2DEG and a
high-frequency electromagnetic field drastically suppresses the
scattering of dressed electrons \cite{Kibis, Morina}. Since the
spin relaxation depends on both the mechanism of spin-orbit
interaction and scattering processes, one can expect that the spin
relaxation time is strongly affected to the dressing
electromagnetic field. Although various mechanisms of spin
evolution in 2DEG have been studied in details both theoretically
and experimentally (see, e.g., Refs. \onlinecite{Wu, Glazov,
Leyland,Tahan}), the spin dynamics of electromagnetically dressed
2DEG escaped the attention before. The present study is aimed to
fill partially this gap at the border between spintronics and
quantum optics.

\section{The spin Hamiltonian of dressed 2DEG} 
\begin{figure}
\includegraphics[width=0.47\textwidth]{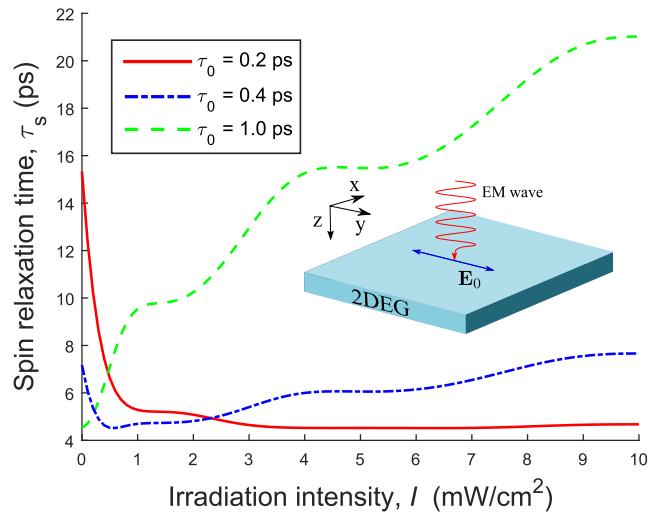}
\caption{(Color online) The dependence of the spin relaxation time
in a 2DEG on the intensity of a dressing electromagnetic field
with the frequency $\omega_0=100$~GHz. The 2DEG is assumed to be
localized in GaAs quantum wells with different initial scattering
times $\tau_0$, the electron effective mass $m=0.067m_0$, the
Fermi energy $\varepsilon_F=10$~meV, and the spin-orbit coupling
constant $\alpha=3.3\times 10^{3}$~m/s. The insert shows the
sketch of the system under consideration.} \label{fig1}
\end{figure}

For
definiteness, we will restrict our consideration to a 2DEG with
the Rashba spin-orbit interaction, which is subjected to a plane
monochromatic linearly polarized electromagnetic wave propagating
perpendicularly to the 2DEG plane (see the insert in
Fig.~\ref{fig1}). In what follows, we will assume that the wave
frequency, $\omega_0$, meets two conditions. Firstly, the wave
frequency is far from resonant electron frequencies corresponding
to interband electron transitions and, therefore, the interband
absorption of the wave by the 2DEG is absent. Secondly, the wave
frequency is high enough in order to satisfy the inequality
$\omega_0\tau_0\gg1$, where $\tau_0$ is the electron scattering
time in an unirradiated 2DEG. It is well-known that the intraband
(collisional) absorption of wave energy by conduction electrons is
negligibly small under this condition (see, e.g.,
Refs.~\onlinecite{Ashcroft,Harrison}). Thus, the considered
electromagnetic wave can be treated as a purely dressing
(nonabsorbable) field. In the absence of scatterers, the wave
function of a dressed electron satisfies the non-stationary
Schr\"odinger equation with the Hamiltonian
\begin{equation}\label{H}
\hat{\cal{H}}=\frac{1}{2m}\left({\hbar\mathbf{k}}-e\mathbf{A}\right)^2+\alpha\left[\bm{\sigma}\times\left({\hbar\mathbf{k}}-e\mathbf{A}\right)\right]_z,
\end{equation}
where ${\mathbf{k}}=({k}_x,{k}_y)$ is the wave vector of the
electron in the 2DEG, $m$ is the effective electron mass in the
2DEG, $e$ is the electron charge,
$\mathbf{A}=({\mathbf{E}_0}/{\omega_0})\cos\omega_0 t$  is the
vector potential of the electromagnetic wave,
$\mathbf{E}_0=(0,E_0,0)$ is the electric field amplitude of the
wave which is assumed to be linearly polarized along the $y$ axis,
$\bm{\sigma}=(\sigma_x,\sigma_y,\sigma_z)$ is the Pauli matrix
vector and $\alpha$ is the Rashba spin-orbit coupling constant. To
simplify calculations, let us subject the Hamiltonian (\ref{H}) to
the unitary transformation
\begin{eqnarray}\label{U}
\hat{U}&=&\frac{1}{\sqrt{2}}e^{i\left(\frac{
k_yeE_0}{m\omega_0^2}\sin\omega_0t-\frac{
e^2E_0^2t}{4m\omega_0^2\hbar}-\frac{
e^2E_0^2}{8m\omega_0^3\hbar}\sin2\omega_0t\right)}\nonumber\\
&\times&
\begin{pmatrix}
e^{i\frac{\alpha eE_0}{\hbar\omega_0^2}\sin\omega_0t}&
e^{-i\frac{\alpha eE_0}{\hbar\omega_0^2}\sin\omega_0t}\\
e^{i\frac{\alpha eE_0}{\hbar\omega_0^2}\sin\omega_0t}&
-e^{-i\frac{\alpha eE_0}{\hbar\omega_0^2}\sin\omega_0t}
\end{pmatrix}.
\end{eqnarray}
Then the transformed Hamiltonian (\ref{H}),
$$\hat{\cal{H}}^\prime=\hat{U}^\dagger\hat{\cal{H}}\hat{U}
-i\hbar\hat{U}^\dagger\frac{\partial U}{\partial t},$$ takes the
form
\begin{equation}\label{H0}
\hat{\cal{H}}^\prime=
\begin{pmatrix}
\hbar^2k^2/2m+\alpha\hbar k_y & -i\alpha \hbar k_xe^{-i\frac{2\alpha eE_0}{\hbar\omega_0^2}\sin\omega_0 t}\\
i\alpha \hbar k_xe^{i\frac{2\alpha
eE_0}{\hbar\omega_0^2}\sin\omega_0 t} & \hbar^2k^2/2m-\alpha \hbar
k_y
\end{pmatrix}.
\end{equation}
Seeking solutions of the Schr\"odinger equation with the
Hamiltonian (\ref{H0}) in the form
\begin{equation}\label{psik}
\psi=
\begin{pmatrix}
a_+\\
a_-
\end{pmatrix}
\end{equation}
and substituting the spinor (\ref{psik}) into the non-stationary
Schr\"odinger equation with the Hamiltonian (\ref{H0}),
$i\hbar\partial\psi/\partial t=\hat{\cal{H}}^\prime\psi$, we
arrive at the system of differential equations
\begin{equation}\label{a}
i\dot{a}_\pm=\left(\frac{\hbar k^2}{2m}\pm\alpha k_y\right)a_\pm
\mp i\alpha k_xa_\mp e^{\mp i\frac{2\alpha
eE_0}{\hbar\omega_0^2}\sin\omega_0 t}.
\end{equation}
Let us apply the Jacobi-Anger expansion \cite{Gradstein},
$$e^{iz\sin\gamma}=\sum_{n=-\infty}^{\infty}J_n(z)e^{in\gamma},$$ to
the exponents in the right side of Eqs.~(\ref{a}) and assume the
2DEG to fill electronic states under the Fermi energy
$\varepsilon_F=\hbar^2k_F^2/2m$. Then Eqs.~(\ref{a}) take the form
which is mathematically equal to the equations of quantum dynamics
of a two-level quantum system under periodical pumping, which are
analyzed in details in conventional textbooks on quantum
mechanics. If the photon energy $\hbar\omega_0$ is much large than
both the Fermi energy $\varepsilon_F$ and the spin-orbit
interaction energy $\alpha\hbar k_F$, the high-frequency harmonics
$e^{in\omega_0 t}$ with $n\neq0$ in the Jacobi-Anger expansion
(``non-resonant terms'') make negligibly small contribution to
solutions of the quantum dynamics equations (\ref{a}) and can be
omitted (see, e.g., the similar analysis for a two-level quantum
system under a periodic pumping in Ref.~\onlinecite{Landau_3}).
Therefore, Eqs.~(\ref{a}) can be rewritten for the considered
high-frequency dressing field as
\begin{equation}\label{a1}
i\dot{a}_\pm=\left(\frac{\hbar k^2}{2m}\pm\alpha
k_y\right)a_\pm\mp i\alpha k_xa_\mp J_0\left( \frac{2\alpha
eE_0}{\hbar\omega_0^2}\right),
\end{equation}
where $J_0(x)$ is the zeros order Bessel function of the first
kind. The equations (\ref{a1}) can be solved trivially and we
arrive at the sought two wave functions (\ref{psik}),
\begin{equation}\label{psi0}
\psi_\pm(\mathbf{k})=
\begin{pmatrix}
\;\;\;\;\;\left[\frac{\sqrt{k_y^2+J^2_0\left(\frac{2\alpha
eE_0}{\hbar\omega_0^2}\right)k_x^2}\pm
k_y}{2\sqrt{k_y^2+J^2_0\left(\frac{2\alpha
eE_0}{\hbar\omega_0^2}\right)k_x^2}}\right]^{\frac{1}{2}}\\
\pm i\left[\frac{\sqrt{k_y^2+J^2_0\left(\frac{2\alpha
eE_0}{\hbar\omega_0^2}\right)k_x^2}\mp
k_y}{2\sqrt{k_y^2+J^2_0\left(\frac{2\alpha
eE_0}{\hbar\omega_0^2}\right)k_x^2}}\right]^{\frac{1}{2}}
\end{pmatrix}
e^{-\frac{i\varepsilon_\pm(\mathbf{k})t}{\hbar}},
\end{equation}
which correspond to the two spin split branches of energy spectrum
of dressed 2DEG,
\begin{equation}\label{En}
\varepsilon_{\pm}(\mathbf{k})=\frac{\hbar^2k^2}{2m}\pm\alpha\hbar
\sqrt{k_y^2+J^2_0\left( \frac{2\alpha
eE_0}{\hbar\omega_0^2}\right)k_x^2}.
\end{equation}

\section{Spin dynamics of dressed 2DEG} In order to analyze
the spin dynamics of the dressed 2DEG under the influence of
scattering processes, let us use a conventional formalism based on
the density matrix response function
\cite{Burkov,Mahan,Burkov1,Mischenko,Stanescu}. The comprehensive
reviews of this theoretical technique can be found, for instance,
in Refs.~\onlinecite{Akkermans, Gumbs, Lerner}. Within this
approach, the evolution of the electron spin
$\mathbf{S}=(S_x,S_y,S_z)$ can be described by the diffusion
equation, $D^{-1}\mathbf{S}=0$, where $D$ is the inverse
propagator of the spin density fluctuation, which is also known as
a diffuson. Assuming the scattering processes in 2DEG to be caused
by a short-range ``white noise'' disorder, the diffuson can be
easily calculated by applying the standard diagram technique
\cite{Burkov,Stanescu}. Writing the diffuson as a sum of single
joint scattering events diagrams $I_{ij}$, we arrive at the
expression $D=(1-I_{ij})^{-1}$. In the case of spatially uniform
electron distribution, the scattering event diagram at the Fermi
level can be expressed in terms of retarded and advanced Green's
functions and is given by
\begin{equation}
I_{ij}=\frac{\hbar}{2\pi\nu_F\tau}\sum_{\mathbf{k}_F}\Tr[G^A({\bf{k}},\varepsilon_F)\sigma_i
G^R({\bf{k}},\varepsilon_F+\hbar\omega)\sigma_j], \label{I}
\end{equation}
where $\nu_F$ is the density of states of 2DEG at the Fermi level,
$\tau$ is the scattering time of 2DEG at the Fermi level, and
$i,j=x,y,z$. Correspondingly, $G^{R(A)}$ in Eq.~(\ref{I}) is the
disorder-averaged single-particle retarded (advanced) Green's
function,
\begin{eqnarray}
G^{R(A)}({\bf {k}},
\varepsilon_F)=\sum_{n=\pm}\frac{\psi_n({\bf{k}})\psi^\dagger_n({\bf{k}})}{\varepsilon_F-\varepsilon_n({\bf{k}})\pm{i\hbar}/{2\tau}},
\label{G}
\end{eqnarray}
which is written in the representation of wave vector $\mathbf{k}$
and frequency $\omega$. Formally, the key expressions (\ref{G})
and (\ref{I}) have the same form for both unirradiated 2DEG and
2DEG subjected to a dressing field. However, for the considered
case of dressed 2DEG, we have to use the wave function of dressed
2DEG (\ref{psi0}) and the energy spectrum of dressed 2DEG
(\ref{En}) in order to calculate the Green's function (\ref{G}).
We have also to take into account that the dressing field
renormalize the scattering time, $\tau$, which takes place both in
Eq.~(\ref{G}) and Eq.~(\ref{I}). Generally, the scattering time is
given by the expression
\begin{equation}\label{tF}
\frac{1}{\tau} =
\sum_{\mathbf{k}^\prime}w_{\mathbf{k}_F\mathbf{k}^\prime},
\end{equation}
where $w_{\mathbf{k}\mathbf{k}^\prime}$ is the electron scattering
probability per unit time between electron states with wave
vectors ${\bf{k}}$ and ${\bf{k}}^\prime$. For the dressed 2DEG,
the scattering probability has the form \cite{Kibis}
\begin{equation}\label{W0}
w_{\mathbf{k}^\prime\mathbf{k}}=J_0^2\left(\frac{e{\mathbf{E}_0}(\mathbf{k}-\mathbf{k}^\prime)}{m\omega_0^2}\right)w_{\mathbf{k}^\prime\mathbf{k}}^{(0)},
\end{equation}
where $w_{\mathbf{k}^\prime\mathbf{k}}^{(0)}$ is the scattering
probability for the 2DEG in the absence of the dressing field.

To simplify calculation of the spin dynamics, let us assume that
the scattering disorder is weak ($\hbar/\tau\varepsilon_F\ll1$)
and the energy of spin-orbit coupling is low ($\alpha\hbar
k_F/\varepsilon_F\ll1$). Performing the integration in
Eq.~(\ref{I}) over the Fermi level, we get matrix elements of the
diffuson. As a final result, we arrive to the spin diffusion
equation $\dot{S}_{x,y,z}=-(1/\tau_{x,y,z})S_{x,y,z}$, where
$\tau_{x,y,z}$ is the sought spin relaxation time for various spin
projections. Since the energy spectrum of dressed 2DEG (\ref{En})
is anisotropic, the spin relaxation times $\tau_x$ and $\tau_y$
are different. However, for realistic parameters of the considered
problem, the spin coupling to the dressing field is very weak
(${2\alpha eE_0}/{\hbar\omega_0^2}\ll1$). Therefore, the
anisotropy of the spin relaxation time in the 2DEG plane can be
neglected and we arrive at the expression
$\tau_{x,y}=\tau_z/2\equiv\tau_s$, where
\begin{equation}\label{taus}
\tau_s=\frac{1+4\zeta^2}{2\zeta^2}\tau
\end{equation}
is the characteristic spin relaxation time in the dressed 2DEG,
$\zeta={\tau}/\tau_{so}$, $\tau$ is the scattering time given by
Eqs.~(\ref{tF})--(\ref{W0}), and $\tau_{so}=1/(\alpha k_F)$ is the
time of spin precession at the Fermi level, which is caused by the
spin-orbit interaction.

\section{Discussion and conclusions} It follows from Eqs.~(\ref{taus})
that the spin relaxation time, $\tau_s$, strongly depends on the
ratio of the scattering time and the spin precession time,
$\zeta={\tau}/\tau_{so}$. Namely, for the case of $\zeta\gg1$, the
spin relaxation time is $\tau_{s}\sim\tau$. On the contrary, for
the case of $\zeta\ll1$, the spin relaxation time is
$\tau_{s}\sim\tau_{so}$. Physically, this strong dependence of the
spin relaxation time (\ref{taus}) on the ratio
$\zeta={\tau}/\tau_{so}$ arises from different mechanisms of spin
relaxation, which are dominant for the cases of $\zeta\gg1$ and
$\zeta\ll1$ (see, e.g., Refs.~\onlinecite{Averkiev, Averkiev1,
Boross}). If the scattering time, $\tau$, is much larger than the
spin precession time, $\tau_{so}$, the spin relaxation is defined
substantially by the scattering processes (the Elliott-Yafet (EY)
spin relaxation mechanism~\cite{Elliott,Yafet}). The EY spin
relaxation alone results in $\tau_{s}\sim\tau$ for the case of
$\zeta\gg1$. If the scattering time, $\tau$, is much less than the
spin precession time, $\tau_{so}$, the spin relaxation is defined
substantially by the spin-orbit interaction (the D'yakonov-Perel'
(DP) spin relaxation mechanism~\cite{D'yakonov}). The DP spin
relaxation alone results in $\tau_{s}\sim\tau_{so}$ for the case
of $\zeta\ll1$. As a consequence, the nonmonotonic dependence of
the spin relaxation time $\tau_{s}$ on the ratio $\tau/\tau_{so}$
appears (see Fig.~2). For an unirradiated 2DEG, the scattering
time $\tau=\tau_0$ depends only on the properties of the given
nanostructure and cannot be easily changed in experiments
(experimentally measured values of the scattering time $\tau_0$ in
various two-dimensional systems can be found, e.g., in
Ref.~\onlinecite{Ando}). On the contrary, in the considered case
of dressed 2DEG, the scattering time $\tau$ depends on both the
dressing field amplitude $E_0$ and the dressing field frequency
$\omega_0$ [see Eqs.~(\ref{tF})--(\ref{W0})] and can strongly
differ from the initial scattering time in unirradiated 2DEG,
$\tau_0$. Therefore, changing the parameters of dressing field, we
can change the value of the scattering time $\tau$. As a result,
the attractive possibility to switch the spin relaxation process
between EY and DP regimes with a high-frequency electromagnetic
field appears.
To clarify the results of numerical calculations of the spin
relaxation time $\tau_s$ [see Figs.~1--2], let us discuss the
dependence of the scattering time (\ref{tF}) on the intensity of
the dressing field $I=\epsilon_0E_0^2c/2$. It follows from the
scattering probability (\ref{W0}) that the dependence arises from
the Bessel function which decreases with increasing the intensity,
$I$. Therefore, the scattering time in 2DEG, $\tau$, increases
with increasing intensity of the dressing field
\cite{Kibis,Morina}. If the initial scattering time in
unirradiated 2DEG, $\tau_0$, is large enough (the dashed line in
Fig.~1), the EY spin relaxation is dominant in the absence of the
dressing field. In this case, the field-induced increase of
scattering time, $\tau$, does not change qualitatively the EY spin
relaxation mechanism. As a result, the relaxation time marked by
the dashed line in Fig.~1 increases monotonically with increasing
the dressing field intensity. On the contrary, if the scattering
time in unirradiated 2DEG, $\tau_0$, is small enough (the solid
and dot-dashed lines in Fig.~1), the DP spin relaxation is
dominant in the absence of the dressing field. In this case, the
field-induced increasing of scattering time, $\tau$, switches the
DP spin relaxation mechanism to the EY one. As a consequence, the
relaxation times marked by the solid and dot-dashed lines in
Fig.~1 demonstrate nonmonotonical behavior with increasing the
dressing field intensity. Therefore, the dressing field can switch
the spin relaxation between DP and EY regimes in a 2DEG with
strong scattering (see the insert in Fig.~2). As to the weak
oscillating behavior of curves in Fig.~1, it is caused formally by
the oscillating behavior of the Bessel function in the scattering
probability (\ref{W0}).
\begin{figure}[h!]
\includegraphics[width=0.47\textwidth]{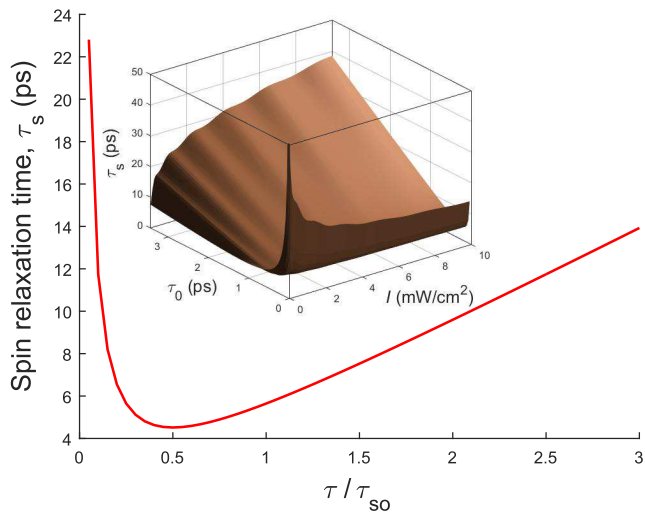}
\caption{(Color online) The dependence of the spin relaxation time
on the ratio of the scattering time in a dressed 2DEG, $\tau$, and
the time of spin precession, $\tau_{so}$, for the 2DEG in a GaAs
quantum well with the electron effective mass $m=0.067m_0$, the
Fermi energy $\varepsilon_F=10$~meV, and the spin-orbit coupling
constant $\alpha=3.3\times 10^{3}$~m/s. The insert demonstrates
the dependence of the spin relaxation time, $\tau_s$, on the
dressing field intensity, $I$, and the initial scattering time,
$\tau_0$, for a dressing electromagnetic field with the frequency
$\omega_0=100$~GHz.} \label{fig2}
\end{figure}

Summarizing the aforesaid, we can conclude that the dressing field
can switch the spin relaxation mechanism in the 2DEG between the
cases corresponding to the well-known Elliott-Yafet and
D'yakonov-Perel' regimes. As a result, the spin properties of the
2DEG can be tuned by a high-frequency electromagnetic field.
Particularly, we showed that the irradiation of 2DEG by the
dressing field results in increasing the spin relaxation time.
Currently, only low-frequency (particularly, stationary) magnetic
and electric fields were considered as a tool to control spin
properties of solids. Therefore, the present effect opens an
alternative way for the spin control with light and, therefore,
forms physical prerequisites for creating light-tuned spintronics
devices.

\begin{acknowledgements}
The work was partially supported by FP7 IRSES projects POLATER and
QOCaN, FP7 ITN project NOTEDEV, Rannis project BOFEHYSS, RFBR
project 14-02-00033, and the Russian Ministry of Education and
Science. We thank D. Yudin for the valuable discussion.
\end{acknowledgements}

\end{document}